\RequirePackage{fix-cm}
\documentclass[twocolumn]{article}

\usepackage[a4paper, margin=0.5in]{geometry}
\usepackage{graphicx}
\usepackage{calc}
\usepackage[table]{xcolor}
\usepackage{authblk}

\def\keywords#1{\\\\\textbf{Keywords}. #1.}
\begin{document}

\title{Crawling Twitter data through API: A technical/legal perspective}

\author[1]{Shahab Saquib Sohail}
\author[1]{Mohammad Muzammil Khan\footnote{Contact: mmkhan.sch@jamiahamdard.ac.in}}
\author[1]{Mohd Arsalan}
\author[2]{\\Aslam Khan}
\author[2]{Jamshed Siddiqui}
\author[3]{Syed Hamid Hasan}
\author[1]{M. Afshar Alam}

\affil[1]{Dept. of CSE, SEST, Jamia Hamdard (India)}
\affil[2]{Dept. of CS, Aligarh Muslim University (India)}
\affil[3]{Fac. of CIT, King Abdulaziz University (Saudi Arabia)}

\date{}

\maketitle

\begin{abstract}
	The popularity of the online media-driven social network relation is proven in today’s digital era. The many challenges that these emergence has created include a huge growing network of social relations, and the large amount of data which is continuously been generated via the different platform of social networking sites, viz. Facebook, Twitter, LinkedIn, Instagram, etc. These data are Personally Identifiable Information (PII) of the users which are also publicly available for some platform, and others allow with some restricted permission to download it for research purposes. The users’ accessible data help in providing with better recommendation services to users, however, the PII can be used to embezzle the users and cause severe detriment to them. Hence, it is crucial to maintain the users’ privacy while providing their PII accessible for various services. Therefore, it is a burning issue to come up with an approach that can help the users in getting better recommendation services without their privacy being harmed. In this paper, a framework is suggested for the same. Further, how data through Twitter API can be crawled and used has been extensively discussed. In addition to this, various security and legal perspectives regarding PII while crawling the data is highlighted. We believe the presented approach in this paper can serve as a benchmark for future research in the field of data privacy.
\end{abstract}

\section{Introduction}
The popularity of the online media-driven social network relation is proven in today’s digital era. The many challenges that these emergence has created include a huge growing network of social relations and the large amount of data which is continuously been generated via the different platforms of social networking sites, viz. Facebook, Twitter, LinkedIn, Instagram, etc. Some of these service providers make data available only for authorized users whereas others provide it publicly. 

The different Social Networking Sites (SNS), viz. Twitter allows downloading their users’ data with some restricted permissions using the Application Programming Interfaces (APIs) [1]. These APIs are an enhanced version of the software that behaves as a back-end interface to collect data from respective SNS. Some of the SNS also provide the impressions of user-praxis by their APIs. The data of the users which are available publicly or through downloading with the help of APIs can be digitally traced [2]. Moreover, with simple or few complex affective computing operations, various personality-traits of the users, their behaviour and social preferences can be explored, which in turn can be misused for manipulating their future actions [3], [4]. Also, pre-defined choices for their future activities can be predicted [5]–[8]. The user, in this case, can be a victim of targeted advertising, Cambridge Analytica like a tragedy, i.e. their preferences can be manipulated, and threat intelligence, etc. 

On the other hand, these data can be very useful for recommendation technology, especially for e-commerce sites and also helps in suggesting users when they are in need of recommendation of medical or psychological treatments. One more important aspect of making the users data public is that the publicly available data of a verified account may ensure security that may arise from fake accounts or unknown malicious handles which need to be monitored and traced.  

Now the problem is how to define that common minimum criterion which can differentiate between the legitimate users and the users which must be monitored? Both the aspects are important enough to be considered but we cannot achieve one on the cost of compromising the other. Therefore, in this paper, we have put up open challenges before the scientific communities related to obtaining the users' data through Twitter API. A detailed procedure for obtaining these users’ data is also illustrated. We have crawled 30 million Tweets during 8 days of continuous data crawling using the Search API. Further, a detailed discussion regarding how these data are prone to various threats is made. We have suggested a framework which considers the fine shades of meaning to appraise the need of preserving users’ privacy as well as avoiding any hindrance to recommender systems so that users can get the benefit from these technologies for the fulfilment of their needs. We believe the suggested framework can work as a benchmark for scientists who use the Twitter APIs for their research and simultaneously for privacy-aware recommender system.

\section{Crawling data through APIs}

\subsection{Background}
The APIs provided by the Social Network Sites (SNS) produce varied and versatile data which can be used by the researchers for their diverse field of researches. These data are users’ personal information, their usage patterns, the network of social connection, et cetera. As far as users’ usage patterns are considered, the data which is collected through APIs are regarded to be more potential and competent than the data which are obtained using interviews or surveys [1], [2], [9]. Few researchers have used Facebook and Twitter supported APIs [10]–[16]. Facebook provides Facebook Query Language as one of the major platforms to download the data of its users. The other APIs include REST API and Facebook Graph API [1], [2]. But Facebook has implemented several restrictions on their APIs. For the purpose of focusing on the central idea as discussed in the previous section, we are confined to Twitter API only.   

It is reported that Twitter, because of its flexibility in providing a degree of freedom to access their database, is the target of most API-centric researches [17]. However, with time, Twitter also has increased the restrictions. Within the first three years, after Twitter was launched, Huberman et al. [18] crawled $3*10^5$ data of Twitter handles to analyze the log and patterns of the users. They also investigated whether a particular tweet is part of the conversation between different Twitter handles or one-way communication?  

H Kwak et al. [19] has crawled almost all the users’ data of that time. They had crawled the data and performed an extensive experiment. Their analyses indicated that almost 85\% of tweets are containing contents related to News. In the same year, the authors in [20] gave the ways to identify the spammers. Their algorithm has a precision of 70\%. Bollier [21] has also suggested how to analyze the data obtained via these social network sites and proposed ideas to conduct quantitative research with a huge amount of data.  

In 2011, Lomborg’s [22] works helped the researchers in identifying several communicative practices over Twitter including social relations and conversational structures, etc. With the help of API, the author gathered different textual information from real communication like; message URLs, IDs of the users, timestamps, etc. In 2012, an open-source Twitter crawler named, TwitterEcho has been developed [23]. They used the REST API for crawling the data. They have also proposed an algorithm to enhance the accuracy of the crawled data. 

Although these works have been reported in the literature regarding the use of APIs for crawling data from Twitter for different purposes, no adequate amount of works have indicated the challenges to the security aspects while obtaining these public data, from the perspectives of users as well as social network sites, both. In the later sections, we have outlined these challenges and proposed solutions to address the issue. 

\subsection{Methodology of crawling Twitter data }
The Crawler uses HTTP GET with application credentials generated from the Twitter Dev Console to request data in JSON from Twitter using the Search API. The rate limit for this API is 450 requests per 15 minutes, with a maximum of 100 tweet count per request. After receiving results from Twitter, the Crawler then parses the JSON into a vector of a custom data type. It saves data in a local directory named “data” and then the date on which it was crawled then the hour, making it structured as “./data/current date/tweets-hour.txt” like “./data/09-07-2019/tweets-20 PM.txt”. The format to save tweet is “CreationDate $<8>$ ID $<8>$ LanguageCode $<8>$ Location $<8>$ Name $<8>$ Username $<8>$ Tweet” delimiter being “$<8>$“. Reason to not use a semicolon (;) or a comma (,) was that a tweet may contain them and processor wouldn’t be able to specify the boundary between entities. The Crawler excluded tweets from users who didn’t enter their location in their profile and tweets in which language wasn’t detected. The Crawler crawled for every 500ms for around 70 hours but after analyzing these data, we found that there were duplicate records. This could be because Twitter systems didn’t differentiate timing in milliseconds or, simply, the data wasn’t being generated that quickly and to fill up the 100 records per request, the API returned duplicates. However, the combination of an interval of 2 seconds and using an API-provided “next” field, which contained a link for the next page in results, gave unique records of tweets. The whole process of analyzing tweets generated by users is divided into 4 parts –  Crawling, Analyzing, Processing, and Pruning. 

\paragraph{Crawling} The crawler uses HTTP GET with application credentials (generated from the Twitter Dev Console) to request data in JSON from Twitter. It uses the Standard (free) Search API provided. The rate limit for this API is 450 requests per 15 minutes, with a maximum of 100 tweet count per request. After receiving results from Twitter, the Crawler then parses the JSON into a local slice of a custom data type. Then the Crawler visits each element of that slice and processes it to remove tweets from which either location wasn’t set by a user or the language didn’t get detected and then it saves it in a special format. If it is an ``original" tweet, it will prepend ``OT" to tweet text otherwise for ``retweets", the API prepends ``RT". It crawled for every 500ms for around 70 hours but the API returned duplicate tweets, for some unknown reason which could be that Twitter systems didn’t differentiate timing in milliseconds. After that, the Crawler crawled every 2 seconds using the ``next" field returned by the API which leads to the next page of tweets and so on. The format to save tweet is ``CreationDate $<8>$ ID $<8>$ Lang $<8>$ Location $<8>$ Name $<8>$ Username $<8>$ Tweet" delimiter being ``$<8>$". Reason to not use a semicolon (;) or a comma (,) was that a tweet may contain them and processor wouldn’t be able to specify the boundary between entities. It saves data in a local directory named ``data" and then the date on which it was crawled then the hour, making it structured as ``./data/current date/tweets-hour.txt" like ``./data/09-07-2019/tweets-20 PM.txt". 

\paragraph{Processing} The code for the processor is written in JavaScript (Node.js). The processor takes the crawled tweets by the crawler and splits the data of individual files by the newline character (``\textbackslash n") and then each line by the custom delimiter ($<8>$). It detects the country and city from the location field. It used an existing detector which usually has low accuracy. The Processor outputs data in the form of JSON and saves it in a file with a structured name ``./data/date-tweets-time.json" like ``09-08-2019-tweets-06 AM.json".

\paragraph{Analysing} The process of analysing is done using the analyser. The code for the analyser is written in ‘GO’ language. The analyzer takes input from the data processed by the Processor and analyzes tweets one by one. It analyzes hard-coded primitive data but can take dynamic Regular Expressions from a file that can be specified by ``-regex" flag. The Analyzer opens a file, loads it in memory, iterates through it and saves any data that matches a description given either by RegEx or hard-coded information. It generates comma-separated-value files (*.csv) for each analysis. 

\begin{figure*}
	\includegraphics[width=1\textwidth]{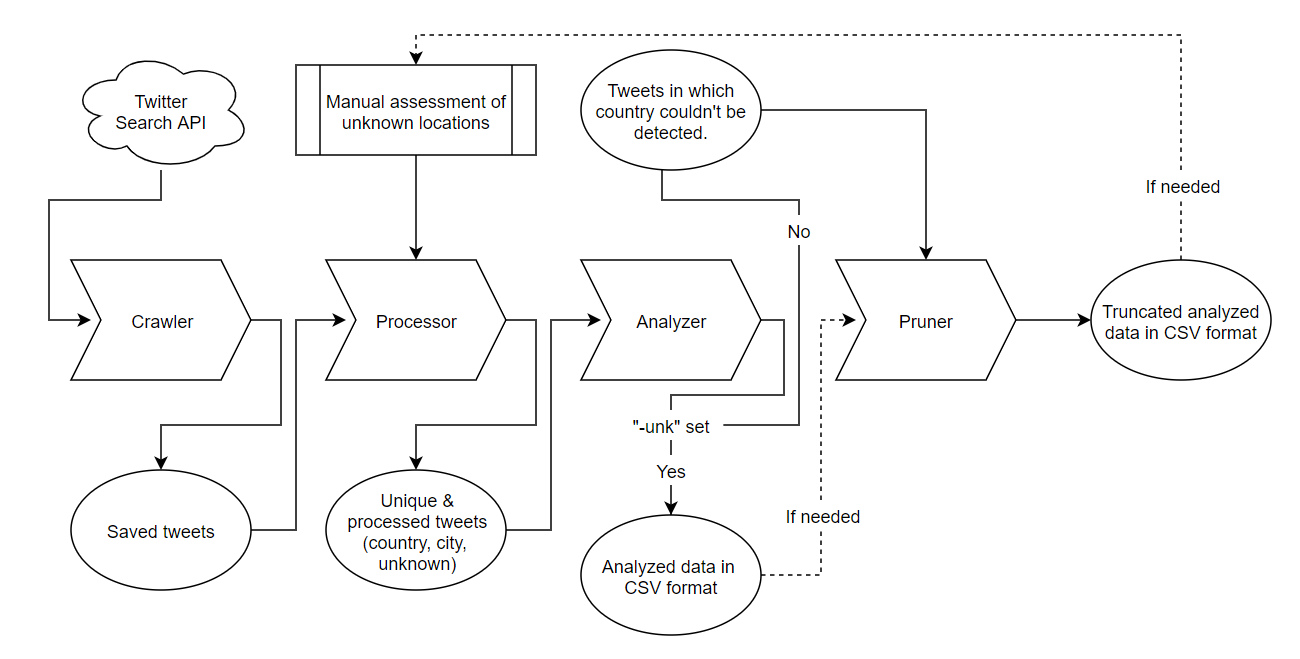}
	\caption{Data Crawling methodology}
	\label{fig:methodology}
\end{figure*}

\paragraph{Pruning} The code for pruner is written in ‘GO’ language Data generated by the Analyzer can be a lot to process. So, Pruner cuts down to relevant information by sorting out the CSV and limiting the number of entries. The workflow is shown in Figure \ref{fig:methodology}.

\section{Security Aspects and Challenges}
\begin{figure*}
	\includegraphics[width=1\textwidth]{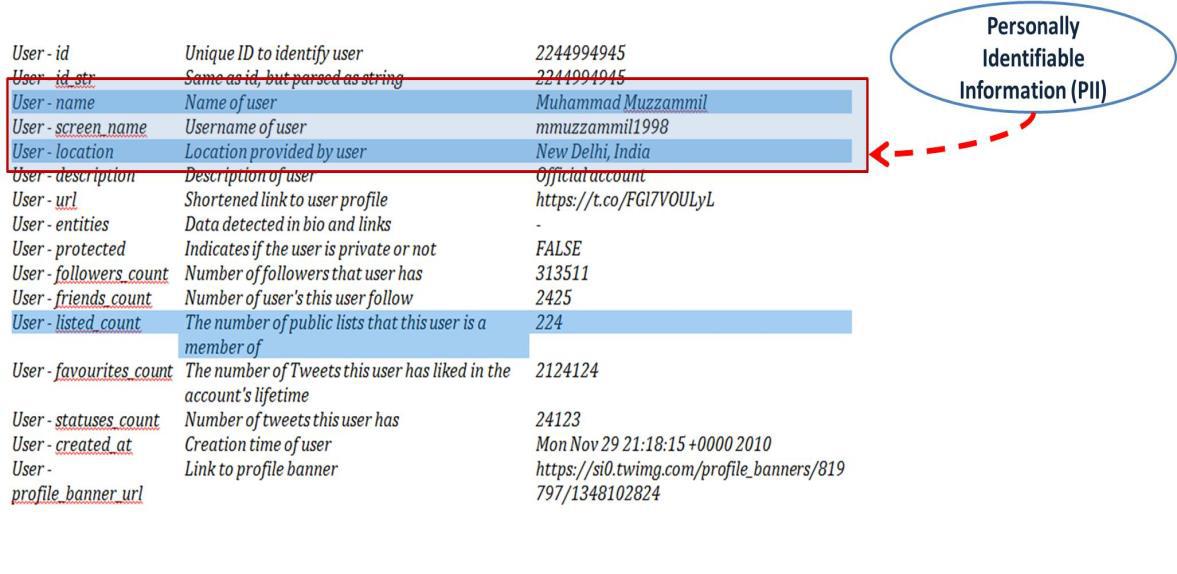}
	\caption{The instance of the feature of the data which has been crawled using Twitter API}
	\label{fig:data}
\end{figure*}

In the previous section data crawling process and description of data has been discussed. How prone to the privacy of an individual these data can be and how much threat it can lead to can be discerned by the advent of Cambridge Analytica (CA) [6], [11]. The exploration of insight in the leaking of personally Identifiable Information by Facebook to CA has poised the attraction of rethinking the access of users’ privacy and has fueled the sensitivity of the impact of technological advancement and social network growth to an individual’s freedom and privacy in particular and society in general. The instance of the feature of the data which has been crawled using Twitter API is shown in Figure \ref{fig:data}.

\begin{figure*}
	\includegraphics[width=1\textwidth]{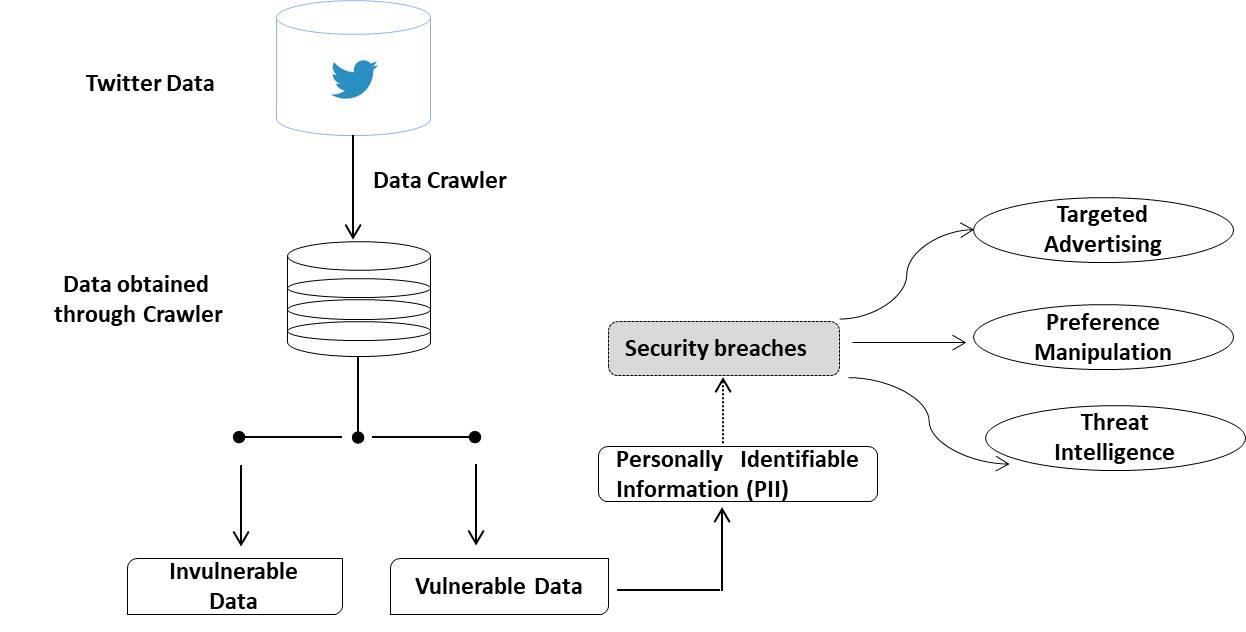}
	\caption{Perspectives of security breaching for data obtained through Crawler using Twitter APIs}
	\label{fig:breach}
\end{figure*}

There are the data which are general in nature and usually not treated as any threat to users. We refer these data as, ‘non-threatening or invulnerable data’. On the other hand, there are the instances of data which can lead to severe threat, which are Personally Identifiable Information of the users and we call it, ‘vulnerable data’. Various studies have been reported in the literature that indicates how crucial is to have a considerable debate on the effects of enhanced technologies to users’ security and their privacy rights [13]. Thus the crawled data can be classified in vulnerable and invulnerable data. The vulnerable data can lead to misuse or unauthorized use of users’ personally identifiable information (PII) resulting in sensitive security breaches. These security breaches can be used for a) threat intelligence b) targeted advertising and c) preference manipulation, etc. Figure \ref{fig:breach} represents the phenomena diagrammatically.

\subsection{Threat Intelligence}
According to a report the cost for security breaches has raised from US\$491 billion in 2014 to US\$1.2 trillion by 2020 [24]. Although the authors have argued that the issue of identifying actionable threat intelligence and their different parameters are yet to be well understood in the literature, we infer that threat intelligence refers to a situation when information about several aspects of users, by any means of intelligence viz. computational, logical, analytical, et cetera are exposed and can create any menace to their privacy, security or personal interests. The threat intelligence, interchangeably used with cyber threat intelligence (CTI), helps in boosting preventive capabilities to better understand the threat information. On the one hand, it makes the user aware of what security threat they may have, on the other hand, actionable threat intelligence indicates an agent who may breach the security can create a menace to the users. With the advanced sentiment analysis algorithm and development of the affective computing, the various users’ personality traits can be predicted [3], [4], [25]–[27]. Thus, the users' online behaviour over social network sites leaves an impression which can suggest someone about their behaviour, thinking and maybe there future actions, which can lead to threat for them. The intelligent machine learning programs and algorithm can decode the users’ plan of actions; hence they can be trapped, misguided or forged for the interest of the atrocious agents. The authors in [28] have proposed an intelligent actionable threat intelligence system. They emphasized on threat management and generating threat information that can arise from any security threats. With the free accessible data of the Twitter handles, they are a prime target of the malicious attacks and they, with any intelligent algorithm, can be trapped due to the expose of either their PII or tweets which in turn can reveal their personal behaviour. 

\subsection{Targeted Advertising (TA)}
The access of the users’ Personally Identifiable Information (PII) can malignly trap them in targeted advertising [29]. With the help of targeted advertising, the users’ data can be exploited for the interest of the agent who is intentionally targeting the users. In an infamous and one of the largest data breaching case in the history of the modern technological era, executed by Cambridge Analytica [6], one of their employees C. Wylie says, “We exploited Facebook to harvest millions of people profiles. And built models to exploit what we know about them and target their inner domain that was the basis the entire company was built on.” The Cambridge Analytica firm took the users personal information without prior permission and started profiling US voters with the help of the system they built for targeting individual voters with political advertisements. These advertisements are personalized and designed for each user that best matches their personalities so that it sounds attractive to them. Most importantly user perceives it as a useful recommendation and never knew that these are targeted ads which were meant to influence their voting [10]. 

The Facebook Platform (FB) has been questioned and was asked for its possible role in targeting voters in the presidential election at US [30]. Since FB and other social network sites need to improve the experience for users. They are allowed to collect user’s friends’ data, however, they are not authorized to sell it or use it for advertisements. Similarly, the Twitter API grants access with restricted permission and prior consent of using it for research and educational purposes only. The availability of these data to someone with a bad intention can cause severe threat which can lead to users’ dissatisfaction, insecurity and violation of users’ privacy policy [31], [32]. Moreover, targeted advertising can be used to make the voters victim by sending them misinformation or open texts from the Internet by combining them with real news without letting them aware the fact that they are targeted political messages meant to misguide the voters. Also, the targeted advertising can be a harmful tool to cheat and forge the innocent people for online shopping and subscribing to different financial schemes, et cetera [7]. 

\subsection{Preference manipulation}
As a consequence of TA, the users can be a victim of preferences manipulation without being aware of it. The preference manipulation includes targeting voters to predict and influence their preferences for the possible candidates, hence the result through ballot boxes. For example, the program can be designed in a way either to identify the individuals may be enticed to vote for their client or maybe discouraged to vote for their opponents even it can play a crucial role in tipping the final result, hence, it can be understood how severe it could be. Further, this manipulation has been observed in the 2016 US election. However, if not the main reason for the outcome, it has enough significance. This degree of significance has been confirmed with the help of research conducted in Stanford on 3500000 users [13].  

Further, the users are menaced with a key approach of being impelled in such a way that they tend to change their behaviour and preferences in the way the concealed targeting agents are guiding to. The preference manipulation mechanism can lead to a grave threat to users including misguiding to opt an object which you never intended for, leaving the most suitable object to least one, and opting for a harmful option without being aware of the fact that the user has been targeted and has become a victim of targeted advertising. How much influence the targeted advertising may have on preference manipulation can be understood by the Cambridge Analytica effects [6] on EU referendum and US election. It is reported [11], [13] that the outcome has been manipulated influenced by CA using targeted advertising.

\section{Privacy protected revelation of Personally Identifiable Information (PII) to third parties}
Summarizing the previous sections, we infer that the public data which are accessed through Twitter API of the users are inherently in conflict with privacy. The major issues are shown in Fig.4. We aimed at providing some kind of technical and legal solutions for the above problems. 

\begin{figure}
	\includegraphics[scale=0.70]{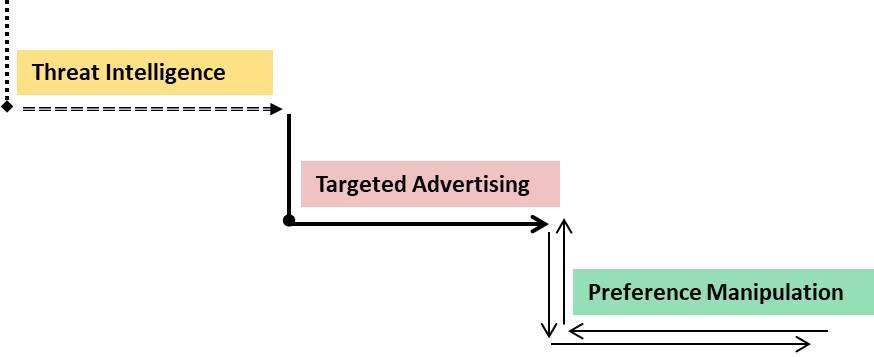}
	\caption{3-steps staircase model for the exploration of security threat to users }
	\label{fig:staircase}
\end{figure}

\subsection{Technical solutions to users’ privacy issues}
\begin{figure*}
	\includegraphics{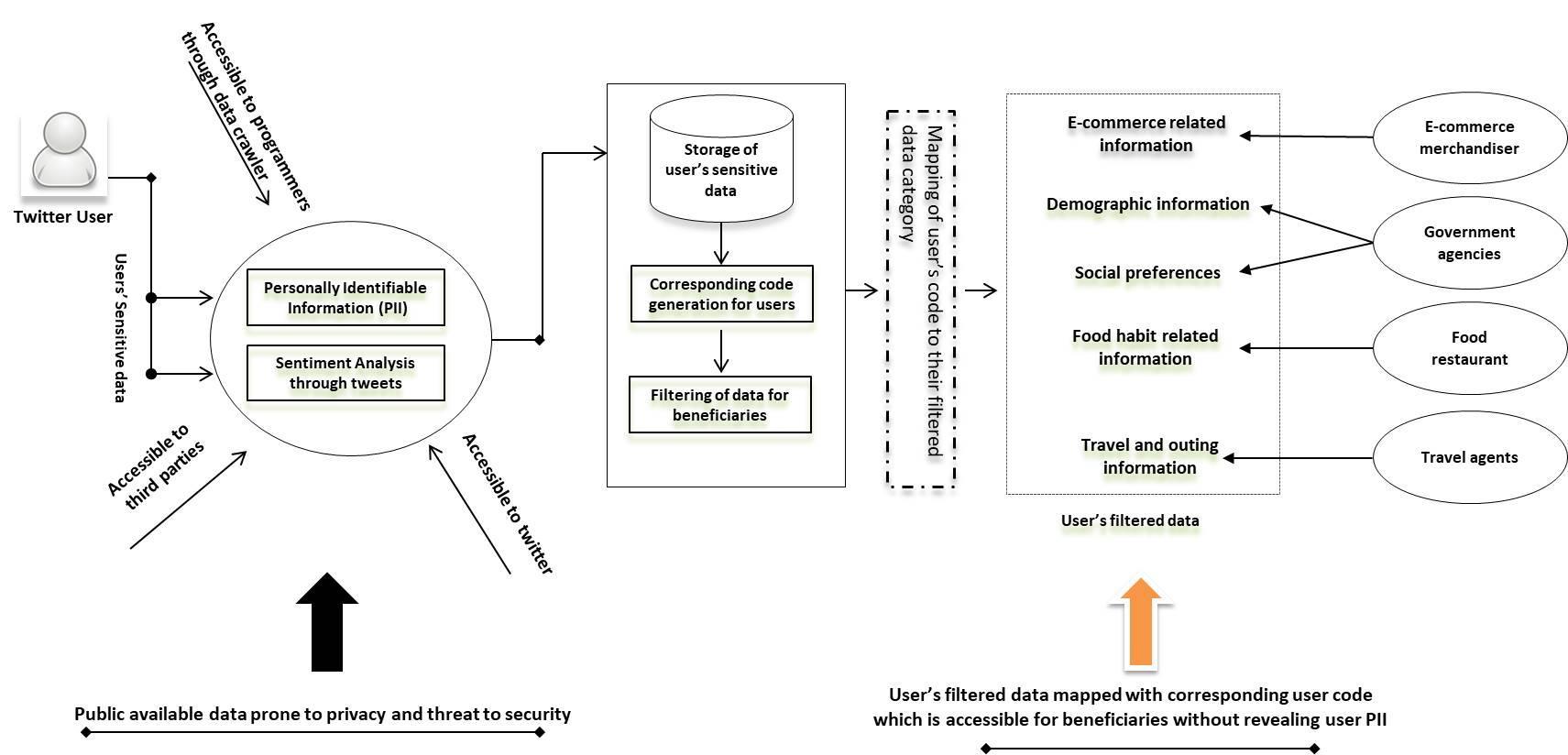}
	\caption{Architecture for privacy-aware recommendation to retain the users’ privacy and PII secure}
	\label{fig:solution}
\end{figure*}

To solve the users’ problems of getting a recommendation for the objects of their tastes, i.e. friends, books [33], [34], items for e-shopping, travel destinations, etc. and keeping their data protected from any embezzling from third parties for their interests to prevent the users from any detriment, a framework has been proposed here. The framework is diagrammatically shown in Fig 5. It has been discussed in the previous sections that with the help of data crawling through APIs users PII can be accessed. Further, the publicly available tweets can expose various personality traits of the users through sentiment analysis [4], [15], [27], [35]. Also, these details help in understanding those aspects of the users which can serve as the base for recommendation technology and providing users with adequate instant resources online. So the need for such a system is crucial [36] which affects neither the users’ online facility nor is the privacy of the user questioned [37], [38]. Therefore, we have proposed a framework which takes both the aspects into consideration. We suggest that public data of the users can be stored and a code is generated corresponding to each user containing the same details but the actual user is hidden from the third parties/software. Moreover, these data with the help of affective computing is classified into respective categories to which the data belong with respect to the recommendation services. These services include e-commerce related information, demographic details and social preferences exploration for mapping to friends, followers of the relatives or suggestion to whom you can follow, food habit, and travel information et cetera, which are usually provided by e-commerce merchandisers, social media, online food ordering Apps, and tour and travel agencies respectively. Thus, the third-party services access the classified information according to their application and need. This information is filtered and transmitted to respective application services. In addition, the corresponding code of the users gets the exact recommendation from the application service providers, which in turn sends back it to the user by re-mapping the code to its owner. By this way, both the purposes are fulfilled without being any threat generated to the users. 

\subsection{Legal aspects of the solutions to users’ privacy issues}
\begin{enumerate}
	\item In addition to technical solutions, we still need clear legal guidance to ensure users’ privacy and data security. For this, by incorporating IEEE-USA digital advice to data safety [40], we have inferred the following suggestions.
	\item Transparency for Public must be made available so that they come to know what data are shared, to whom it has been shared, and for how long they can retain these data with them. Also, wherever and whenever the third parties use their data, they must have prior permission from the legitimate owner of the data or at least the users must be informed if the transfer of the data is not going to harm and are considered to be least sensitive. Further, it should also be made compulsory for the agents or third parties to make it public for the users to know the mechanism by which data is being captured. The process needs to be as easy as an average user can identify and understand it.
	\item The information at different web sites or any third party repositories should also be disclosed to users. And all the beneficiaries from the data must also be notified to users. iii) Any information which is based on users’ personally identifiable information (PII) must be designed in such a way that users whenever they would like to eliminate the information or delete the text from the existing location, can do. Further, if they wish to remove their PII, the system should be flexible enough that they can remove these data. 
	\item Any loss of the private data of the users must be intimidated to them. and any data related to minor users must also be specified.  
\end{enumerate}

\section{Conclusion}
In this paper, a comprehensive data crawling mechanism is presented, moreover, the legal and technical aspects regarding privacy and security issues concerned with the users and data both are discussed. We have also concluded that the deciding line in making data public is the common minimum criterion which can differentiate between the legitimate users and the fake users. At the same time neither we can compromise users’ privacy nor can we expose to fake users. Hence, an open question to scientific communities regarding access of personally identifiable information (PII) of the users through Twitter API has been put up, and technical as well as a legal model for the solution has also been proposed. The proposed solutions can serve as a benchmark for future researches. In future, it is recommended to evaluate the proposed model to boost the mechanism.

\end{document}